\documentclass[onecolumn]{elsart3p}

\newcommand{\Jnatphys}{Nat. Phys.}

\newcommand{\Jscience}{Science}

\newcommand{\Jrmp}{Rev. Mod. Phys.}

\newcommand{\crasphy}{C. R. Phys.}

\newcommand{\Jprocroysoc}{Proc. Roy. Soc. A: Math. Phys. Eng. Sci.}

\newcommand{\Jijthphys}{Int. J. Theor. Phys.}

\newcommand{\Jstatphys}{J. Stat. Phys.}

\newcommand{\JRepProgPhys}{Rep. Prog. Phys.}
\newcommand{\JjphysCM}{J. Phys.: Cond. Matt.}

\newcommand{\JFortschrPhys}{Fortschr. Phys.}

\usepackage{amssymb}
\usepackage{amsmath}
\usepackage[english,french]{babel}
 \usepackage[T1]{fontenc}

\newtheorem{e-proposition}[theorem]{Proposition}

\newtheorem{e-definition}[theorem]{Definition\rm}

\renewcommand{\theequation}{\arabic{equation}}
\usepackage{graphicx}

\def\og{\leavevmode\raise.3ex\hbox{$\scriptscriptstyle\langle\!\langle$~}}
\def\fg{\leavevmode\raise.3ex\hbox{~$\!\scriptscriptstyle\,\rangle\!\rangle$}}

\usepackage{color}

\newcommand{\ie}{i.e.}
\newcommand{\eg}{e.g.}

\renewcommand{\etal}{\textit{et al.}}

\begin{document}
\centerline{Quantum simulation}
\begin{frontmatter}

\selectlanguage{english}
\title{Quantum simulation: From basic principles to applications}

\selectlanguage{english}

\author[authorlabel1]{Laurent Sanchez-Palencia}
\ead{lsp@cpht.polytechnique.fr}
\address[authorlabel1]{CPHT, Ecole Polytechnique, CNRS, Universit\'e Paris-Saclay, Route de Saclay, 91128 Palaiseau cedex, France}

\end{frontmatter}

\selectlanguage{english}


\bigskip

Envisioned by Richard Feynman in the early 1980s, quantum simulation has received dramatic impetus thanks to the development of a variety of plateforms able to emulate a wide class of quantum Hamiltonians. During the past decade, most
of the quantum simulators have implemented rather well-known models, hence permitting a direct comparison with theoretical calculations and a precise benchmarking of their reliability. The field has now reached a maturity such that one can
address difficult problems, which cannot be solved efficiently using classical algorithms. These advances provide unprecedented opportunities to explore previously unreachable fields, test theoretical predictions, and inspire novel approaches.

This contribution is an elementary introduction to quantum simulation. We discuss the challenges, define both digital and analog
quantum simulators, and list the demanding conditions they require. We also provide a brief account of the contributions
gathered in the dossier on \textit{Quantum Simulation} of the Comptes-Rendus de Physique of the French Academy of Sciences~\cite{tarruell2018,aidelsburger2018,lebreuilly2018,LeHur2018,bell2018,alet2018}.
The latter completes excellent reviews that appeared previously, see for instance Refs.~ \cite{buluta2009,georgescu2014,NaturePhysicsInsight2012cirac,NaturePhysicsInsight2012bloch,NaturePhysicsInsight2012blatt,NaturePhysicsInsight2012aspuru-guzik,NaturePhysicsInsight2012houck,ward2013}.

\bigskip
\noindent{\bf Universal models and the role of simulations in many-body physics} \\
Understanding the behavior of macroscopic quantum systems is a major challenge of modern physics. The basic laws of
low-energy physics are by now quite well known at the microscopic, say atomic, scale. Conversely, many fundamental
questions remain open, and even debated, about the collective dynamics at the macroscopic scale. By "macroscopic scale",
here we mean systems made up of a huge number of constituents, or degrees of freedom, say $10^6$, $10^{23}$, or even more, as
relevant in condensed matter physics or in astrophysics, for instance. Such huge systems cannot be treated exactly, be it a
the classical level and, even worse, at the quantum level. Yet, it is the main outcome of the thermodynamic approach that
the collective behavior of a macroscopic system can drastically differ from that of its elementary constituents. For instance,
the elementary interactions between the $\textrm{H}_2\textrm{O}$ molecules are fundamentally unchanged when a water bucket turns from the
liquid phase to the solid phase at zero degree Celsius. Similarly, the interactions between the microscopic magnetic moments
do not show any brutal change when a magnetic material gets magnetized underneath the Curie temperature. Hence, the
dramatic effects observed in macroscopic systems are governed by large-scale instabilities, without obvious counterparts at
the microscopic level. This observation takes a universal character, summarized in the celebrated motto "More is Different"~\cite{anderson1972}.
Such so-called emerging phenomena also appear in quantum systems, where new effects that are impossible in the
classical world show up below some critical temperature or at zero temperature when some interaction parameter passes
through a critical value. Celebrated examples include the superfluid $\Lambda$ transition in helium or super-conducting transitions
and other metal-insulator transitions in electronic systems.

Strikingly enough at first sight, while emerging phenomena only appear in very large scale macroscopic systems, their
germs are contained in the mutual interactions of their elementary constituents. Hence, local two-body interactions are
sufficient to explain a huge class of phase transitions, such as the liquid-solid and magnetic transitions mentioned above.
The Boltzmann statistical approach proved particularly successful in describing this connection between the microscopic
and macroscopic scales. To understand emerging phenomena on grounds as fundamental as possible, the programme is
now well established: One tries to identify the basic microscopic terms that seem to be relevant and disregards all the
other microscopic details. One then elaborates a model, as generic as possible, likely to reproduce the main experimental
observations. The most usual examples are the Ising and Heisenberg models for magnetic transitions, or the Hubbard model
for metal-insulator transitions~\cite{mahan2000,bruus2004,tsvelik2007}.
Then all is left to do is to check that the phenomenon of interest indeed emerges
from the dynamics of the simplified model. The realization of this programme is nothing but a simulation. It consists in
building up a simplified system that mimics the main properties of a real system.

\bigskip
\noindent{\bf The difficulty of simulating generic quantum systems using classical computers} \\
Unfortunately, severe problems usually appear at the last stage because limiting ourselves to a simple model does not guarantee that the solution is straightforward. In a few cases, exact solutions are known, a significant class of which is Bethe-ansatz
integrable models. In some other cases, it is possible to make relevant approximations and build up tractable theories that
yield accurate predictions. This is the case of mean field approximations, which work well to explain some superfluid transitions for instance. In most cases, however, exact or quasi-exact solutions are not known. One then traditionally resorts
to numerical simulations. In this respect, the development of advanced approaches, such as Monte Carlo techniques, density functional theory, molecular dynamics, tensor-network approaches, and dynamical mean field theory, to name a few,
have dramatically contributed to enlarging the class of models whose solution is known. However, even the most advanced
numerical approaches have their own limitations and use approximations or representations that do not hold in all cases.
Then, it is necessary to turn to exact computations. While it may be possible for some reasonably large classical systems, it
is practically impossible for a quantum system.

Consider the most simple example put forward by Richard Feynman in Ref.~\cite{feynman1982} of a system made of $N$ spins $1/2$.
Since each spin can be either in the spin-up or spin-down state, there are $2^N$ possible configurations.
A pure classical state is parametrized by $N$ binary numbers, the values of which, $0$ or $1$, represent the state of each spin.
A computer with a memory of $100\,\textrm{Go} \sim10^{12}\,\textrm{bits}$ can thus efficiently simulate about $10^{12}$ spins $1/2$.
Conversely, a generic pure quantum state is the coherent superposition of all the possible configurations. This requires $2^N$ $\mathbb{C}$-coefficients and the same memory of $100\,\textrm{Go}$ can only store the spin state of only $\log(10^{12})/\log(2) \sim 40$ particles.
It is thus impossible to simulate the exact
quantum state of more than a few tens of spins. More generally, it is practically impossible to store and manipulate the
state of a macroscopic quantum system, owing to the exponential growth of its Hilbert space in the system's physical size.

A careful reader might note that a similar issue occurs if one wants to simulate the full statistical distribution of the
classical counterpart of the same spin systems. In this case, one would need to store the probabilities of each of the $2^N$ configurations, the number of which also grows exponentially with the number of constituents, $N$. However, this issue can be easily
solved by using a stochastic algorithm, which amounts to introduce random jumps between the configurations. This is what
Monte Carlo algorithms do for instance. Then, relevant quantities may be found by running the simulation a large, but not
exponentially large, number of times, and averaging the results. As noticed by Feynman, this just simulates what Nature
indeed does when we acquire experimental data~\cite{feynman1982}.

The issue is much more serious in the quantum world, because even pure states cannot be simulated efficiently on truly
large scales. Then, Feynman pointed out that the only reasonable thing to do is to make the computer quantum itself. Then,
the register would naturally be exponentially large in the number of bits and one would be able to store an exponentially
large number of coefficients~\cite{feynman1982}. Feynman also postulated that it should be possible to simulate the time evolution of any
quantum system in a reasonable computer time. This statement was proven for local quantum Hamiltonians a few years
later~\cite{lloyd1996}.
Seith Lloyd showed that the simulation can be performed with arbitrary precision in a computer time that grows
at most polynomially with the physical time, using Trotter-Suzuki's decomposition of the many-body evolution operator into
sequences of local Hamiltonian evolutions. A machine, exploiting quantum resources to simulate a given quantum model, is
called a \textit{quantum simulator}.

Up to now, we have focused on the dynamics of many-body systems at thermodynamic equilibrium. However, similar
questions and challenges appear when one considers the unitary time evolution of large systems. Realizing quantum simulations in this case is actually even more important for several reasons. On the one hand, we do not know (yet) a universal
approach, which would be the out-of-equilibrium counterpart of what is statistical physics for equilibrium thermodynamics.
It follows that many questions are still completely open, regarding for instance the spreading of information or the thermalization of these systems. On the other hand, out-of-equilibrium quantum systems are likely to develop an entanglement
entropy that grows unbounded during the time evolution. It results that simulations using classical machines rapidly reach
their limits.

\bigskip
\noindent{\bf What is a quantum simulator?} \\
Building up a quantum simulator is still a \textit{tour de force}. Of course, the aim here is not to reproduce exactly the
initial system, for instance a complete material with all its microscopic details. Should we do this, we would loop back to the
initial problem and we would gain no information about the relevance of the theoretical model. In the best case, we would
address the specific initial problem and miss the basic physical ingredients at the origin of the considered phenomenon.
Instead, the aim here is to build up a clean system, exactly governed by a basic model Hamiltonian. The latter is proposed
in a theoretical context to reproduce some physical phenomenon observed in a class of real systems. This is the basic idea
of quantum simulation.

More precisely, a quantum simulator is a controlled device that allows us to
(i)~engineer a class of quantum Hamiltonians exactly,
(ii)~control its dynamics, and
(iii)~make sufficiently precise measurements to consider that the problem is solved.
In practice, a quantum simulator is useful provided it is able to solve a hard problem. In this context, one considers that
a problem is "hard" when it cannot be solved, either analytically or numerically with a classical computer in a time that
grows at most as a power of the system's size. This feature is often retained as the main criterion for a quantum simulator. However, it cannot be considered a perennial definition owing to continuous progress in many-body analytical and
numerical approaches. In fact, it is very hard to prove that a given quantum problem cannot be solved with a classical algorithm in polynomial time. However, it is a common belief that a large amount of entanglement is a good working criterion.
Such a situation appears, for instance, in the vicinity of quantum phase transitions, in gapless systems at equilibrium or
in far-from-equilibrium systems where the entanglement may grow unbounded during the time evolution. The applications
of quantum simulation are potentially unlimited and range from condensed-matter and high-energy physics to chemistry,
biology, and cosmology, for instance~\cite{buluta2009,georgescu2014,NaturePhysicsInsight2012}.

\smallskip
\noindent
To be more specific, one distinguishes two classes of quantum simulators:

\smallskip
\noindent
\textit{Digital quantum simulators~--~}
A digital quantum simulator is a universal machine, fully reprogrammable to simulate the
thermodynamics or the real-time evolution of any quantum model. This is the kind of simulators envisioned by Feynman and Lloyd. It was proven that, at least for local Hamiltonians, \ie, Hamiltonians containing only finite-range interactions, such
a universal quantum simulator can indeed be implemented~\cite{feynman1982,lloyd1996}. Such a machine would exploit a fully reconfigurable
register of qubits and a programmable sequence of logical gates to realize the desired simulation. This is actually nothing
but a quantum computer~\cite{benioff1980,deutsch1985,divincenzo1995,divincenzo2000}.
Hence, the distinction between a digital quantum simulator and a quantum computer
mainly lies on the use we make of it. While a "quantum computer" would be used to implement a variety of quantum
algorithms, the most celebrated example of which are the Shor factorization algorithm or the Grover search algorithm, a
"universal quantum simulator" would be dedicated to optimization problems, such as the determination of the ground state
of some Hamiltonian, or to its unitary, real-time, dynamics. In principle, building a universal quantum simulator would thus
be as difficult as building a quantum computer. In particular, it would be sensitive to the same decoherence issues and
require the implementation of error-correction codes. However, real or imaginary time evolution by a local Hamiltonian
usually requires much less resources than generic quantum algorithms, and, in practice, are significantly more robust.

\smallskip
\noindent
\textit{Analog quantum simulators~--~}
An alternative approach consists in building up a physical system from scratch to simulate
each specific model. Assume one wants to determine the ground state or the time evolution of some Hamiltonian $\hat{H}$, say a spin-$1/2$ model.
The idea behind analog quantum simulation is to create an ensemble of elements with two well-identified
states, \eg, the two polarizations of a photon, the two internal states of an atom, or a quantum dot. These two states are
used to represent the two spin states. Then isolate the system from its environment and engineer interactions between these
two-state elements according to the Hamiltonian $\hat{H}$. This may be realized by coupling the system with a cavity in the case
of photons or via laser fields in the case of atoms, for instance. In order to find the ground state, cool down the system;
in order to determine its time evolution, prepare it in a well-defined initial state and let it run. Then, one realizes from
scratch a system exactly governed by the desired model Hamiltonian and Nature simply works for us, with all its quantum
properties. The last thing to be done is to measure the outcome.

Analog quantum simulation has pros and cons compared to digital quantum simulation. On the one hand, one needs to build up a new analog
simulator for each studied model, while a unique digital simulator would be sufficient. On the other hand, the architecture
of the analog simulator can be optimized for the considered problem, while that of a digital simulator is the result of
a compromise between all possible cases. In practice, an analog quantum simulator is much easier to build than a digital
quantum simulator and there are now many examples of successful implementations of the former, see for instance Ref.~\cite{NaturePhysicsInsight2012} and the contributions in this special issue~\cite{tarruell2018,aidelsburger2018,lebreuilly2018,LeHur2018,bell2018,alet2018}.

\bigskip
\noindent{\bf Requirements and challenges}\\
By analogy with standard numerical simulation, one may say that digital quantum simulation assumes that the hardware
is available and focuses on the software part. Conversely, analog quantum simulation works directly on the former and the
software part is built on the latter. In both cases, realizing an efficient quantum simulator requires to take up several major
challenges that we summarize here (see also Ref.~\cite{NaturePhysicsInsight2012cirac}):
\begin{itemize}\label{QSrequirements}
\item[(i)] \textit{Build up.} Create a quantum system that can be manipulated almost at will using external fields. It should be isolated
from its environment to avoid decoherence issues. Depending on the problem at hand, the system can be made of
bosons, fermions, spins, or mixtures of the latter.
\item[(ii)] \textit{Quantum engineering.} Design the desired Hamiltonian with at least one adjustable relevant parameter. One should be
able to tune the latter from a regime where the problem can be solved by other means to a regime where it cannot be
solved easily. It allows benchmarking of the quantum simulator, similarly as it is done in traditional numerical work.
\item[(iii)] \textit{Initialization.} Prepare the system in a well-defined state. It allows one to target the ground state using cooling techniques or to prepare some initial state to explore a specific trajectory in the system's Hilbert space, for instance. The
initial state is often a pure state, but one can also prepare a mixed state by interaction with a controlled bath. Furthermore, the bath may be used for simulating open quantum systems.
\item[(iv)] \textit{Detection.} Measure relevant, local or non-local, observables that yield sufficient information to "solve" the problem with sufficient fidelity. Depending on the problem at hand, one may use destructive or non-destructive measurement techniques.
\end{itemize}
Since the goal is to use quantum simulators to solve problems than cannot be solved by other means, a major concern
is of course the reliability of the used simulator. There are several ways to address this issue~\cite{hauke2012}.
First, as mentioned above, one can make sufficient benchmarking of the simulator by addressing regimes where other solutions exists. Second,
in the case of isolated systems, one can check that quantum coherence is maintained using an adiabaticity property: one back-propagates the system and checks the fidelity of the final state to the initial state. Third, it is crucial to solve a
given problem on several simulators based on significantly different platforms. Should the results be consistent, one would
consider that the solution is reliable. In this respect, dramatic progress has been realized starting by the early 2000s in a
variety of fields, including:
\begin{itemize}
\item[(*)] Ultracold quantum gases,
\item[(*)] Artificial ion crystals,
\item[(*)] Photonic systems, including polaritons in cavities,
\item[(*)] Superconducting circuits,
\item[(*)] Magnetic insultors,
\item[(*)] Electronic spins in quantum dots,
\end{itemize}
for instance. During the past decade, a number of quantum simulators have been demonstrated. So far, the results of most of
them could be directly compared to theoretical calculations, which allowed extensive benchmarking. Now, some of the most
advanced implementations are likely to address really hard problems, which cannot be solved efficiently using classical
algorithms. To make this exciting perspective a reality, continued development of a variety of platforms, including those
mentioned above and hopefully new ones, will be pivotal. It will allow us to address complementary questions as well as
to compare results obtained by different approaches.

\bigskip
\noindent{\bf Contributions to this dossier}\\
The dossier on \textit{Quantum simulation} makes a point on recent advances of the field and discusses perspectives via a selection
of contributions in various areas from ultracold atoms and quantum optics to statistical physics and condensed matter.
Tarruell and myself review progress on the quantum simulation of the celebrated Hubbard model using ultracold Fermi gases.
Aidelsburger~\etal provide an introduction to novel approaches to engineer artificial gauge fields within a wide class of systems ranging from quantum optics to solid-state systems.
Lebreuilly and Carusotto discuss realizations of strongly correlated
quantum fluids of light in driven-dissipative photonic devices with applications to the generation of Mott insulator and fractional quantum Hall states of light.
Le Hur~\etal review advances in the study of real-time dynamics of impurity models and
their realizations in quantum devices, including superconducting circuits, quantum electrical circuits, and ultracold-atom architectures.
Bell~\etal propose a novel platform based on superconducting quantum interference devices (SQUIDs) to emulate quantum phase transitions in one dimension, as well as perspectives to address non-integrable and disordered systems.
Finally, Alet and Laflorencie discuss recent advances on many-body localization in isolated quantum systems and current experimental efforts to probing this physics.

\bigskip
\noindent{\bf Acknowledgment}\\
I am grateful to Jean~Dalibard and Daniel~Estève for suggesting the dossier on \textit{Quantum Simulation} and to Christophe Salomon for tacking over the editorial process of the contribution I am an author of~\cite{tarruell2018}.
I also thank Jean~Dalibard for useful comments on this manuscript.

\bibliographystyle{naturemag}


\end{document}